\documentclass[a4paper,fleqn,usenatbib]{mnras}

\usepackage[T1]{fontenc}
\usepackage{ae,aecompl}

\usepackage{graphicx}	
\usepackage{amsmath}	
\usepackage{amssymb}	
\usepackage{adjustbox}
\usepackage{xspace}
\usepackage{natbib}
\usepackage{etoolbox}
\makeatletter
\patchcmd\@combinedblfloats{\box\@outputbox}{\unvbox\@outputbox}{}{%
  \errmessage{\noexpand\@combinedblfloats could not be patched}%
}%
\makeatother
\usepackage{cleveref}
\crefname{figure}{Figure}{Figures}
\crefname{section}{Section}{Sections}
\crefname{table}{Table}{Tables}

\newcommand{\Msol}{M$_{\odot}$\xspace}

\newcommand{\squotes}[1]{\lq {#1}\rq\xspace}
\newcommand{\dquotes}[1]{\lq\lq {#1}\rq\rq\xspace}
\newcommand{\eagle}{{\sc eagle}\xspace}
\newcommand{\illustris}{{\sc illustris}\xspace}
\newcommand{\gadget}{{\sc gadget}\xspace}
\newcommand{\subfind}{{\sc subfind}\xspace}
\newcommand{\anarchy}{{\sc anarchy}\xspace}
\newcommand{\galform}{{\sc galform}\xspace}
\newcommand{\LAGN}{$L_{\mathrm{AGN}}$\xspace}
\newcommand{\LXray}{$L_{\mathrm{2-8keV}}$\xspace}
\newcommand{\Msolyr}{M$_{\odot}$~yr$^{-1}$\xspace}
\newcommand{\SFR}{$\dot M_{*}$\xspace}
\newcommand{\BHAR}{$\dot M_{\mathrm{BH}}$\xspace}

\newcommand{\M}[1]{$M_{\mathrm{#1}}$\xspace}
\newcommand{\av}[1]{$\langle$#1$\rangle$\xspace}

\title[Galaxy and BH growth in the \eagle simulation]{The link between galaxy and black hole
growth in the \eagle simulation}

\author[S. McAlpine et al.]{Stuart McAlpine,$^{1}$\thanks{E-mail: s.r.mcalpine@durham.ac.uk}
Richard G. Bower,$^{1}$
Chris M. Harrison,$^{2}$
Robert A. Crain,$^{3}$
\newauthor
Matthieu Schaller,$^{1}$
Joop Schaye$^{4}$
and Tom Theuns$^{1}$
\\
$^{1}$Institute for Computational Cosmology, Department of Physics, University of Durham, South Road, Durham, DH1 3LE, UK\\
$^{2}$Centre for Extragalactic Astronomy, Department of Physics, Durham University, South Road, Durham DH1 3LE, UK\\
$^{3}$Astrophysics Research Institute, Liverpool John Moores
  University, 146 Brownlow Hill, Liverpool L3 5RF, UK\\
$^{4}$Leiden Observatory, Leiden University, P.O. Box 9513, 2300 RA
  Leiden, the Netherlands
}

\date{Accepted XXX. Received YYY; in original form ZZZ}

\pubyear{2016}

\begin{document}
\label{firstpage}
\pagerange{\pageref{firstpage}--\pageref{lastpage}}
\maketitle

\begin{abstract}

We investigate the connection between the star formation rate (SFR) of galaxies
and their central black hole accretion rate (BHAR) using the \eagle
cosmological hydrodynamical simulation.  We find, in striking concurrence with
recent observational studies, that the \av{SFR}--BHAR relation for an AGN
selected sample produces a relatively flat trend, whilst the \av{BHAR}--SFR
relation for a SFR selected sample yields an approximately linear trend.  These
trends remain consistent with their instantaneous equivalents even when both
SFR and BHAR are time-averaged over a period of 100~Myr.  There is no universal
relationship between the two growth rates. Instead, SFR and BHAR evolve through
distinct paths that depend strongly on the mass of the host dark matter halo.
The galaxies hosted by haloes of mass \M{200} $\lesssim 10^{11.5}$\Msol grow
steadily, yet black holes (BHs) in these systems hardly grow, yielding a lack
of correlation between SFR and BHAR. As haloes grow through the mass range
$10^{11.5} \lesssim$ \M{200} $\lesssim 10^{12.5 }$\Msol BHs undergo a rapid
phase of non-linear growth. These systems yield a highly non-linear correlation
between the SFR and BHAR, which are non-causally connected via the mass of the
host halo.  In massive haloes (\M{200} $\gtrsim 10^{12.5}$\Msol) both SFR and
BHAR decline on average with a roughly constant scaling of SFR/BHAR $\sim
10^{3}$.  Given the complexity of the full SFR--BHAR plane built from multiple
behaviours, and from the large dynamic range of BHARs, we find the primary
driver of the different observed trends in the \av{SFR}--BHAR and
\av{BHAR}--SFR relationships are due to sampling considerably different regions
of this plane.  

\end{abstract}


\section{Introduction}
\label{sect:introduction}

Substantial effort has been dedicated both observationally and theoretically to
identifying the link between the growth of galaxies and their central
supermassive black holes (BHs). However, the nature of this relationship
remains poorly understood.  Indirect evidence of a causal connection has been
suggested empirically based on the \textit{integrated} properties of galaxies
and their BH counterparts. For example, galaxy bulge mass (\M{*, bulge}) and
the mass of the central BH (\M{BH}) exhibit a tight, approximately linear
correlation for bulge masses in excess of \M{*, bulge} $\sim 10^{10}$\Msol
\citep[e.g,][]{Magorrian1998,Kormendy2013,McConnellandMa2013,Scott2013}.
However, at lower bulge mass, a steeper trend has been advocated
\citep[e.g,][]{Scott2013,Greene2016}.  Additionally, the cosmic star formation
rate (SFR) and black hole accretion rate (BHAR) densities broadly trace one
another through time
\citep[e.g,][]{Heckman2004,Aird2010,MadauandDickinson2014}.

A simple interpretation for these global relationships is that the growth rates
that build these properties (i.e. the SFR of the galaxy and accretion rate of
the BH) are proportional throughout their evolution, thus growing the two
components in concert. More complex evolutionary scenarios have also been
proposed. For example, a simple time-averaged relationship built from a common
fuel reservoir of cold gas \citep{AlexanderandHickox2012,Hickox2014}, a rapid
build up of galaxy and BH mass via merger induced starburst/quasar activity
\citep[e.g,][]{Sanders1988,DiMatteo2005,Hopkins2008} or a mutual dependence on
the mass or potential of the dark matter halo
\citep{BoothandSchaye2010,BoothandSchaye2011,Bower2017}.  In these scenarios
the SFR and BHAR do not necessarily trace each other directly and the observed
correlations may only appear in massive galaxies due to an averaging of very
different histories.  Furthermore, \citet{Peng2007} and \citet{Jahnke2011} go
as far as to suggest there is no causal connection of any kind, with correlations
only appearing as result of a random walk. 

To test these scenarios, numerous observational studies have attempted to
identify a direct link between the intrinsic growth rates of galaxies and their
central BHs. Studies that investigate the mean SFR (\av{SFR}) as a function of
BHAR consistently find no evidence for a correlation for moderate-luminosity
sources \citep[\LXray $\lesssim$ $10^{44}$~erg~s$^{-1}$;
e.g,][]{Lutz2010,Harrison2012,Page2012,Mullaney2012b,Rosario2012,Stanley2015,Azadi2015}.
For high-luminosity sources (\LXray > $10^{44}$~erg~s) however, there has been
significant disagreement as to if this relation becomes positively correlated
\citep[e.g.][]{Lutz2010}, negatively correlated \citep[e.g.][]{Page2012} or
continues to remain uncorrelated
\citep[e.g,][]{Harrison2012,Rosario2012,Stanley2015,Azadi2015}.  These
disparities between various works at the high-luminosity end are likely due to
small number statistics and sample variance \citep{Harrison2012}, and indeed,
recent studies using large sample sizes confirm the extension of a flat trend
to higher luminosities \citep{Stanley2015,Azadi2015}.

A flat trend for the \av{SFR}--BHAR relation could potentially be interpreted
as revealing an absence of a connection between SFR and BHAR. However, studies
that have investigated the mean BHAR (\av{BHAR}) as a function of SFR
consistently find a \textit{positive} relationship
\citep[e.g,][]{Rafferty2011,Symeonidis2011,Mullaney2012a,Chen2013,Delvecchio2015}.
Within the paradigm of a linear \M{BH}--\M{bulge,*} relation due to a universal
co-evolution of BH and galaxy growth, both approaches are expected to produce a
consistent, similarly linear result (see \cref{sect:motivation} for a
derivation of why this is).  \citet{Hickox2014} proposes a potential solution,
suggesting that SFR and BHAR are connected \textit{on average} over a period of
100~Myr, with a linear scaling. This relationship disappears when measured
instantaneously owing to the rapid variability timescale of AGN, with respect
to that of galactic star formation.  

From a theoretical perspective, many simulations have focussed on the growth of
BHs in galaxy mergers \citep[e.g,][]{DiMatteo2005,Hopkins2005}. Whilst both
star formation and BH accretion are typically enhanced during the merger
proper, the extent of the connection between SFR and BHAR pre- and post-merger
event remains unclear. \citet{Neistein2014} demonstrate through the use of a
semi-analytical model that the observed correlations between galaxies and
their central BHs can be reproduced when BH growth occurs only during merger
induced starbursts. This could explain the lack of a correlation between growth
rates in low-luminosity systems whilst allowing for mutual enhancement during
the merger events themselves.  \citet{Thacker2014} investigate the impact of
various feedback models on the SFR--BHAR parameter space in a set of equal mass
merger simulations. They find a complex evolution for individual systems, even
when averaged over 20~Myrs. Any correlation found is strongly dependant on the
feedback model chosen, with the post-merger phase showing the strongest
evidence for a positive connection. Using a high-resolution hydrodynamical
merger suite, \citet{Volonteri2015a} find BHAR and galaxy-wide SFR to be
typically temporally uncorrelated. They suggest in \citet{Volonteri2015b} that
the observed discrepancy between the \av{SFR}--BHAR and \av{BHAR}--SFR
relations seen observationally are a result of sampling two different
projections of the full bi-variate SFR--BHAR distribution whose build up is
constructed from different behaviours between SFR and BHAR before, during and
after the merger event. 

It is now possible to extend these investigations to within a full cosmological
context. Using the semi-analytical code \galform, \citet{Gutcke2015} find a
negative SFR--AGN luminosity correlation at low AGN luminosities, this then
transitions to a strong positive correlation at high AGN luminosities.  In the
cosmological hydrodynamical simulation \illustris, \citet{Sijacki2015} find a
single trend in the SFR--BHAR relationship embedded in a large scatter,
particularly in BHAR.  Cosmological hydrodynamical simulations have the
advantage of probing the entire galaxy population within a self consistent
variety of environments with a diverse range of accretion and merger histories.
Here we investigate to what extent galaxy and BH growth rates are connected
within the \dquotes{Evolution and Assembly of GaLaxies and their Environment}
\citep[\eagle,][]{Schaye2015,Crain2015} \footnote{\url{www.eaglesim.org}}
\footnote{Galaxy and halo catalogues of the simulation suite are publicly
available at \url{http://www.eaglesim.org/database.php} \citep{McAlpine2015}.}
simulation.  Adopting physical prescriptions for radiative cooling, star
formation, stellar mass loss, BH accretion, BH mergers and both stellar and AGN
feedback, \eagle reproduces many observed properties of galaxies, BHs and the
intergalacic medium with unprecedented fidelity \citep[e.g,][]{Schaye2015,
Furlong2015a, Furlong2015b, Trayford2015, Schaller2015a, Lagos2015,
Rahmati2015, Bahe2016, Crain2016, RosasGuevara2016, Segers2016, Trayford2016}. 

The paper is organised as follows.  In \cref{sect:simulationsandsubgrid} we
provide a brief overview of the \eagle simulation suite, including the subgrid
model prescription and simulation output.  The results are presented in
\cref{sect:results}.  We examine the \eagle predictions of the \av{SFR}--BHAR
relationship for an AGN selected sample and the \av{BHAR}--SFR relationship for
a SFR selected sample in \cref{sect:observations}, finding good agreement to
recent observational findings.  To investigate why these trends might be
different, we explore the effect of time-averaging each growth rate and examine
potential sampling biases in \cref{sect:understanding}.  \cref{sect:to_halo}
examines the influence of the host dark matter halo on both SFR and BHAR,
finding that each exhibits a strong connection.  Finally in
\cref{sect:discussion}, we discuss the changing relationship between SFR and
BHAR as the halo grows and in \cref{sect:conclusions}, we present our
conclusions.

\section{Simulations \& Subgrid model}
\label{sect:simulationsandsubgrid}

\begin{figure*}
\includegraphics[width=\textwidth]{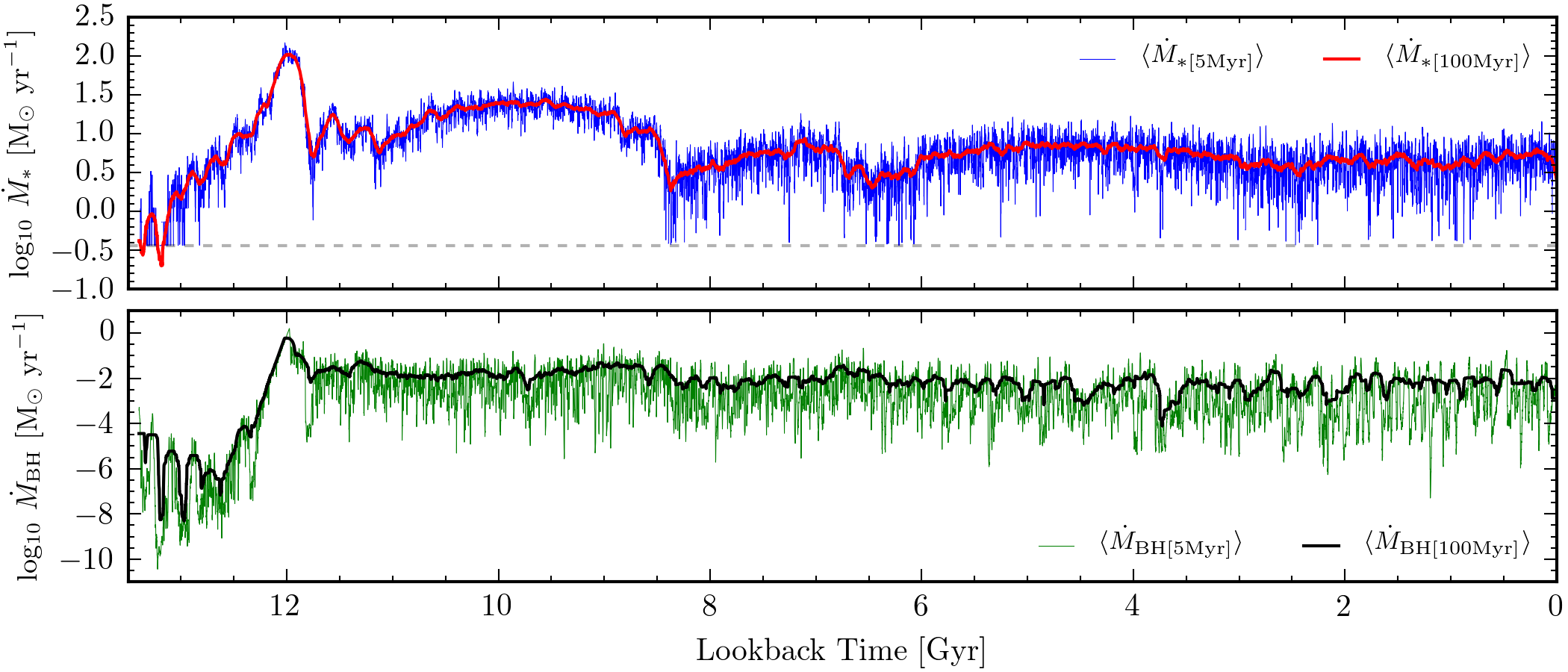}

\caption{Galaxy (top panel) and BH (bottom panel) growth rates as a function of
lookback time for an individual galaxy ({\tt GalaxyID}=20324216,
$M_{\mathrm{200[z=0]}}=10^{13.2}$\Msol, $M_{\mathrm{*[z=0]}}=10^{11.1}$\Msol,
$M_{\mathrm{BH[z=0]}}=10^{8.7}$\Msol). Blue (green) and red (black) lines show
the SFR (BHAR) history averaged over 5~Myr and 100~Myr respectively. When
averaged over short timescales, BHARs can vary by as much as $\approx 4$ dex.
SFRs however vary considerably less over the same time window ($\approx 1$
dex), and generally represent values in much closer agreement to their long
term average rate. In this individual case the long term average trends of SFR
and BHAR yield quite different evolutionary behaviours. As the both the
particle mass and averaging time window is finite, the minimum possible SFR
sampled for a 5~Myr timescale is shown as a dashed grey line.} 

\label{fig:variability}
\end{figure*}

\eagle is a suite of cosmological hydrodynamical simulations comprising a range
of periodic volumes, numerical resolutions and physical models. The simulations
are run using a substantially modified version of the N-body TreePM smoothed
particle hydrodynamics code \gadget-3 \citep{Springel2005}, referred to as
\anarchy \citep[Dalla Vecchia, \textit{in prep}; see also see also appendix A
of][]{Schaye2015}.  For this study we focus on the largest run
(Ref-L0100N1504), a cubic periodic volume of 100 comoving megaparsecs (cMpc) on
each side, containing $1504^{3}$ dark matter particles of mass $9.7 \times
10^{6}$~\Msol and an equal number of baryonic particles with an initial mass of
$1.8 \times 10^{6}$~\Msol. The subgrid parameters are those of the \eagle
reference (\squotes{Ref-}) model, described fully by \citet{Schaye2015}.
Cosmological parameters are those inferred by \citet{Planck2013}, namely:
$\Omega_{\mathrm{m}}=0.307$, $\Omega_{\Lambda}=0.693$,
$\Omega_{\mathrm{b}}=0.04825$, $h = 0.6777$ and $\sigma_{8}=0.8288$.

\subsection{Subgrid model}

Processes operating below the numerical resolution of the simulation are
treated as \squotes{subgrid}, implemented as a series of physical models. A
detailed description of the full subgrid prescription is given by
\citet{Schaye2015}, with consideration to their influence on the reference
model given by \cite{Crain2015}. Here we give only a brief overview:

\begin{itemize}

\item \textit{Radiative cooling and photo-ionisation heating} is implemented as
per \citet{Wiersma2009a}, tracing 11 elements in the presence of the cosmic
microwave background and the evolving, spatially uniform UV/X-ray background of
\citet{HaardtandMadau2001}. 

\item \textit{Star formation} is implemented as a pressure dependent relation
that reproduces the Kennicutt-Schmidt law described by
\citet{Schaye_DallaVecchia2008}. The subsequent \textit{stellar mass loss} via
winds of massive stars and supernovae is computed as per \citet{Wiersma2009b}. 

\item \textit{Stellar feedback} is injected thermally and stochastically
following the method of \citet{DallaVecchia_Schaye2012}.

\item \textit{BH seeding} follows the prescription first introduced by
\citet{Springel2005a}, whereby BHs are introduced as collisionless sink
particles placed in the centres of dark matter haloes more massive than $1.475
\times 10^{10}$\Msol, which do not already contain one. BHs enter the
simulation with a seed mass $m_{\mathrm{seed}} = 1.475 \times 10^{5}$\Msol
and subsequently grow via accretion of surrounding gas or mergers with other
BHs.

\item \textit{BHs grow via accretion} of nearby material at a rate estimated
from the modified Bondi-Hoyle formalism introduced in \citet{RosasGuevara2015}.
In short, the model is an extension of the spherically symmetric case of
\citet{Bondi1944} accounting now for the circularisation velocity of the
surrounding gas, capped at the Eddington limit. Contrary to
\citet{RosasGuevara2015}, we do not use an additional boost factor ($\alpha$).

\item \textit{AGN feedback} is implemented as a single mode, where it is
injected thermally and stochastically into the surrounding interstellar medium
as per \citet{Booth_Schaye2009}. Feedback is performed assuming a single
efficiency, independent of halo mass and accretion rate. 

\end{itemize}

As described by \citet{Crain2015}, the subgrid model parameters are calibrated
to reproduce the observed galaxy stellar mass function, galaxy sizes and
normalisation of the \M{BH}--\M{*,bulge} relation at $z \approx 0.1$.

\subsection{Simulation output}

\subsubsection{Halo and galaxy identification}

Outputs are stored as 29 \squotes{snapshots} between redshifts $z=20$ and $z=0$
at which the complete state of every particle is recorded. In addition, 400
data-lite \squotes{snipshots} are produced, with a typical temporal separation
of $\approx$40-60~Myr. Bound structures are identified in post processing.
First, dark matter haloes are identified using the \squotes{Friends of Friends}
(FOF) algorithm with linking length of $b = 0.2$ times the mean interparticle
separation \citep{Davis1985}.  Then, bound substructures (or
\squotes{subhaloes}) within these haloes are identified with the \subfind
program \citep{Springel2001,Dolag2009} applied to the full particle
distribution (dark matter, gas, stars and BHs). We associate the baryonic
component of each subhalo with a galaxy, defined to be the \textit{central}
galaxy if it hosts the particle with the minimum gravitational potential and
the remainder being classified as \textit{satellites}.

Halo mass, \M{200}, is defined as the total mass enclosed within
$r_{\mathrm{200}}$, the radius at which the mean enclosed density is 200 times
the critical density of the Universe. Galaxy mass, \M{*}, is defined at the
total stellar content belonging to a subhalo within a 30~pkpc spherical
aperture as per \citet{Schaye2015}. 

\subsubsection{Constructing histories of individual galaxies}
\label{sect:individual_galaxies}

In order to accurately trace the evolution of individual galaxies and their
central BH for the analysis in \cref{sect:to_halo}, we require histories of a
higher temporal resolution than is provided by the snipshot output. To do this,
we follow galaxies and their central BH through cosmic time. As a galaxy
descendant may have multiple progenitors, we trace the progenitor galaxy that
is hosted along the \squotes{main progenitor branch} of the merger tree as
defined by \citet{Qu2017}, the branch containing the greatest total mass along
its history.  

BHARs are recorded at each timestep with a typical spacing of $\sim 10^{3} -
10^{4}$~yr, yielding an \squotes{instantaneous} rate. These can then be
time-averaged over longer durations. Quoted \squotes{instantaneous} SFRs are
taken from the snapshot output, where they are computed based on the current
star forming state of the gas contained within the galaxy. Time-averaged SFR
histories are constructed from the stellar particles born within the main
progenitor that reside in the galaxy at the present day. As these particles
store both their birth time and initial mass, collectively they create a robust
history of star formation for that galaxy.  However, as these histories are
sensitive in their resolution to the number of particles sampled, only galaxies
containing more than 200 particles (\M{*[z=0]} $\approx 10^{8.5}$\Msol) are
considered for this study. 

\cref{fig:variability} shows an example history of an individual galaxy's SFR
(top panel) and accretion rate of the central BH (bottom panel) through cosmic
time taken from the methods described above. We show each growth rate
time-averaged over 5~Myr (blue and green lines) and 100~Myr (red and black
lines) to highlight the large difference in variability scatter between the two
timescales. This is particularly severe for BHAR, where values recorded over
short timescales do not return a good approximation of the long term average
rate, differing in value by as much as 4~dex. We have adopted 100~Myr as our
long averaging duration as it reflects an estimate of the effective timescale
for empirical indicators of star formation using the far-infrared \citep[FIR,
the tracer of star formation for the observational studies compared to in
\cref{sect:observations}, see the discussions
by][]{Neistein2014,Volonteri2015a}. Although there are similar features between
SFR and BHAR through time for this individual case (for example a common peak
at a lookback time of 12 Gyr), globally the two histories are quite different. 

\subsection{The \M{BH}--\M{200} relation}
\label{sect:eagle_bhs}

\begin{figure}
\includegraphics[width=\columnwidth]{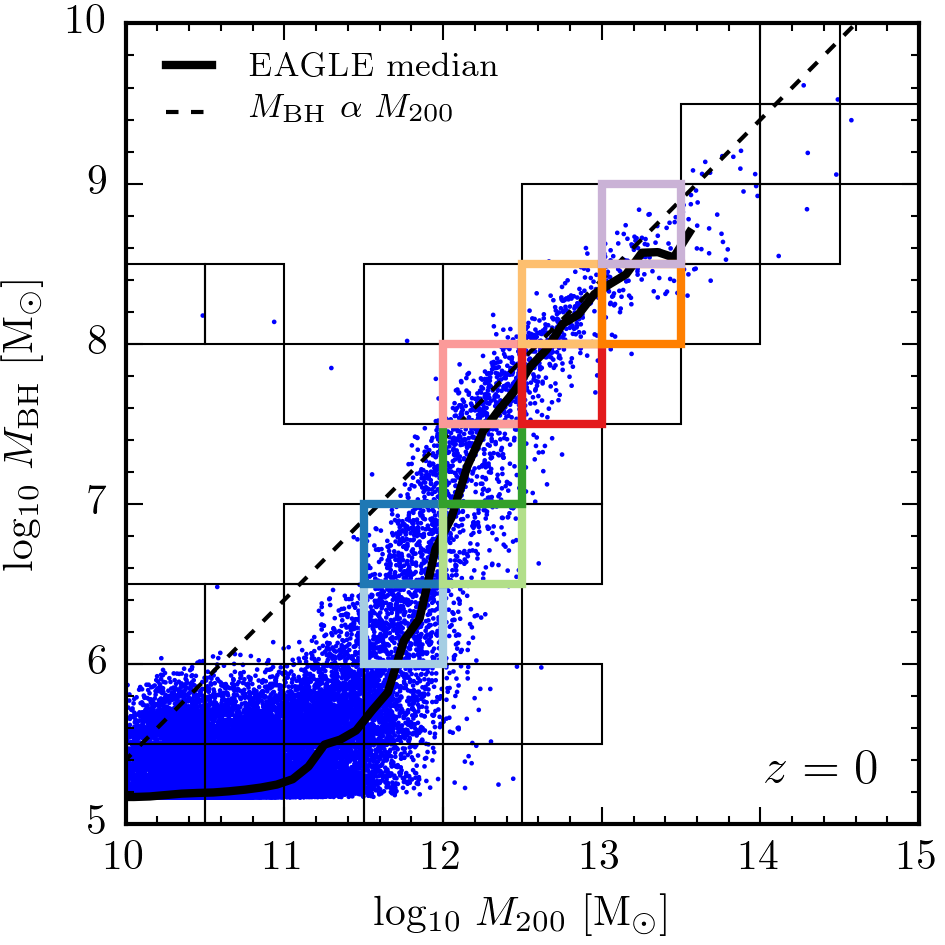}

\caption{\M{BH}--\M{200} relation for central galaxies at $z=0$, where \M{200}
is the halo mass. Each galaxy is represented by individual blue points and the
median trend is shown as a solid black line. The black dashed line shows a
linear relationship, \M{BH} $\propto$ \M{200}, for reference.  Overlaid
two-dimensional bins are 0.5~dex on each side and contain at least one galaxy.
Nine of these cells are used for the continued investigation in
\cref{sect:discussion} and are outlined in colours that relate to the histories
shown in \cref{fig:avHistory_vs_hm,fig:sfr_vs_bhar_av}.}

\label{fig:bhm_vs_hm}
\end{figure}

\cref{fig:bhm_vs_hm} shows the \M{BH}--\M{200} relation for central galaxies at
$z=0$.  We have plotted \M{BH} as a function of halo mass rather than bulge or
total stellar mass due to the crucial connection that \M{200} has with both SFR
and BHAR (see \cref{sect:to_halo} onwards). We note that the \M{BH}--\M{*}
relation also follows the same behaviour \citep[see Figure 1 of][]{Barber2016}
and throughout this description \M{200} and \M{*} can be interchanged. The
overlaid two dimensional bins are for the continued investigation in
\cref{sect:to_halo}, where they are fully described.  

The empirical relationship between BH mass and that of the classical bulge is
well described by a single power law at high mass \citep[e.g,][with gradient
values of $\alpha \approx 1-1.3$ satisfying
\cref{eq:slope1}]{Magorrian1998,Kormendy2013,McConnellandMa2013}.  Indeed, one
of the calibration parameters of the simulation is to match the normalisation
of this relationship.  However, whereas traditionally this trend has been
linearly extrapolated to lower-mass systems, \eagle predicts a steepening of
the trend.  As a consequence, the relation between BH mass and the mass of the
host galaxy or halo is not well described by a single power law. Interestingly,
a steeper slope at intermediate masses is supported by recent observations of
bulge (or pseudobulge) systems \citep[e.g,][]{Scott2013,Greene2016}.  When
total stellar mass is considered as the independent variable,
\citet{Reines2015} predict Seyfert-like systems yield and alternate
\M{BH}--\M{*} relationship to previously measured early-type systems. However,
each trend is consistent with a linear relation, with Seyfert-like systems
harbouring a lower normalisation.

For \eagle, BHs in massive systems (\M{200} $\gtrsim 10^{12.5}$\Msol) follow an
approximately linear trend with halo mass (compare to the black dashed black
line in \cref{fig:bhm_vs_hm}), but those hosted by haloes with mass \M{200}
$\lesssim 10^{12.5}$\Msol follow a much steeper relation and those in the
lowest mass systems (\M{200} $\lesssim 10^{11.5}$\Msol) plateau at the seed BH
mass.

\citet{Bower2017} argue that multiple physical processes drive the
relation between \M{BH} and \M{200}. In low (high) mass systems stellar (AGN)
feedback regulates the baryonic inflow to the galaxy, suppressing BH (continued
stellar) growth.  There is a critical transition halo mass (\M{200} $\sim
10^{12}$\Msol, hereafter \M{crit}) separating these two regulatory regimes.
Within \M{crit} haloes, neither feedback process is dominant, and as a result
BHs grow at a highly non-linear rate. These phases create the flat,
supra-linear and $\sim$linear regimes of BH growth seen in the integrated
quantities of \cref{fig:bhm_vs_hm} and have important consequences for the
galaxy and BH growth rates investigated throughout this study.

\subsection{Absolute calibration of SFRs}
\label{sect:sfr_calibration}

When comparing to the observed cosmic SFR density, \citet{Furlong2015a} found an
almost constant -0.2~dex offset for redshifts $z \leq 3$.  There is however
continued uncertainty as to the absolute calibration of SFR indicators on which
these observations rely.  For example, \citet{Chang2015} find upon revisiting
this calibration with the addition of WISE photometry to the full SDSS
spectroscopic galaxy sample that the SFRs of local galaxies along the
main-sequence are systematically lower than previously estimated by $\approx
0.2$~dex, yielding good agreement with the \eagle prediction \citep[see Figure
5 of][]{Schaller2015b}.

As the observational datasets compared to in \cref{sect:observations} utilise
an earlier calibration, we \textit{reduce} all observed SFRs by 0.2~dex. The
magnitude of this recalibration is shown as a red arrow in
\cref{fig:stanley2015,fig:delvecchio2015}. This serves to remove the known
global systematic offset, making it simpler to focus on the trends with BHAR
that are the topic of this paper. 

\section{Results}
\label{sect:results}

\subsection{Comparison to observations}
\label{sect:observations}

\begin{figure}
\includegraphics[width=\columnwidth]{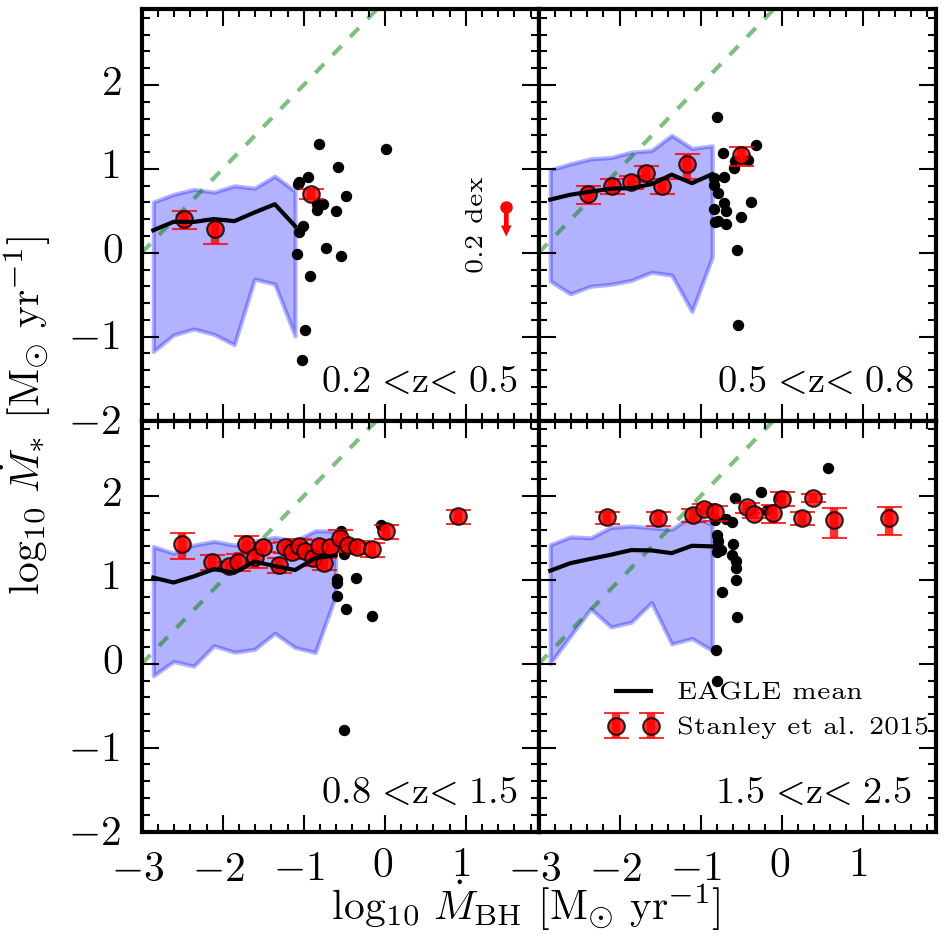}

\caption{\SFR--\BHAR (SFR--BHAR) relation for four continuous redshift ranges
from $0.2 < z < 2.5$. BHARs are instantaneous and SFRs are the time-averaged
rate over the 100~Myr preceding the BH event. The \textit{mean} SFR as a
function of BHAR for central galaxies is shown as a black line, with the
corresponding blue shaded region indicating the 10-90$^{\mathrm{th}}$
percentile range. Bins containing fewer than 10 objects have their galaxies
represented individually as black solid circles.  The linear relation
\BHAR/\SFR = $10^{-3}$ is shown as a dashed green line and the data of
\citet{Stanley2015} is represented as red circles.  Fits to the \eagle mean
relations are tabulated in \cref{table:slopes}. The magnitude of the SFR
recalibration applied to the data for all redshifts is indicated by a red arrow
in the upper left panel (see \cref{sect:sfr_calibration}). For each redshift
range we find mean trends that are considerably flatter than a linear relation
($\gamma_{\mathrm{S15}} \ll 1$ in \cref{eq:slope4}).}

\label{fig:stanley2015}
\end{figure}

\begin{figure}
\includegraphics[width=\columnwidth]{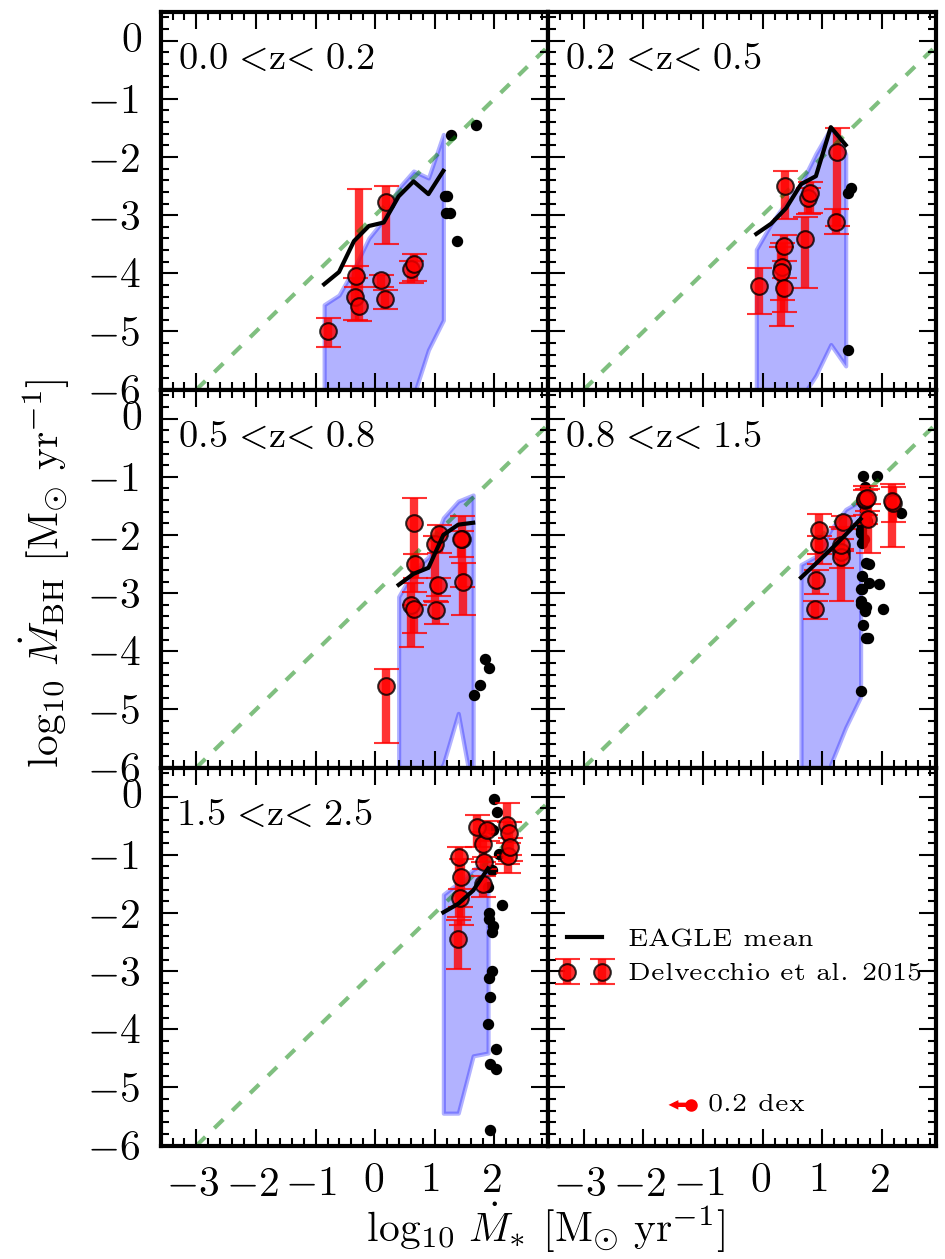}

\caption{\BHAR--\SFR (BHAR--SFR) relation for five continuous redshift ranges
from $0.0 < z < 2.5$. BHARs are instantaneous and SFRs are the time-averaged
rate over the 100~Myr preceding the BH event. The \textit{mean} BHAR as a
function of SFR for central galaxies is shown as a black line, with the
corresponding blue shaded region indicating the 10-90$^{\mathrm{th}}$
percentile range. Bins containing fewer than 10 objects have their galaxies
represented individually as black solid circles.  The linear relation
\BHAR/\SFR = $10^{-3}$ is shown as a dashed green line and the data of
\citet{Delvecchio2015} is represented as red circles. Fits to the \eagle mean
relations are tabulated in \cref{table:slopes}. The magnitude of the SFR
recalibration applied to the data for all redshifts is indicated by a red arrow
in the lower right panel (see \cref{sect:sfr_calibration}). For each redshift
range we find gradients of the mean trend close to unity ($1 /
\gamma_{\mathrm{D15}} \approx 1$), in good agreement with a linear relation.}

\label{fig:delvecchio2015}
\end{figure}

\begin{table*}

\caption{Slope ($\gamma$) and intercept ($\delta$) of various relations
satisfying \cref{eq:slope4}. S15 denotes the study of \citet{Stanley2015} and
D15 denotes the study of \citet{Delvecchio2015}.  Columns 2 and 3 are the
fitted values of the \eagle mean relations investigated in
\cref{fig:stanley2015,fig:delvecchio2015}. Annotated with a * in columns 6 and
7 are the \eagle \emph{median} fits to the same datasets. Fits to the mean
relation of the 100~Myr time-averaged growth rates from \cref{sect:time_av} for
the same datasets are shown in columns 4 and 5. Errors on the individual
mean/median \eagle data points are taken from bootstrap re-sampling. Quoted
fits and their associated errors were computed using the Python module
\sc{lmfit}.}

\begin{tabular}{crrrrrrr} \hline
$z$ &
$\gamma_{\mathrm{S15}}$  &
log$_{10} \delta_{\mathrm{S15}}$  &
$\gamma_{\mathrm{S15}} \langle \mathrm{100 Myr} \rangle$ &
log$_{10} \delta_{\mathrm{S15}} \langle \mathrm{100 Myr} \rangle$ &
$\gamma_{\mathrm{S15}}$*  &
log$_{10} \delta_{\mathrm{S15}}$* \\
\hline\hline
0.20 < $z$ < 0.50 & $0.13 \pm 0.04$ & $0.67 \pm 0.09$ & $0.31 \pm 0.05$ & $1.18 \pm 0.14$ & $0.14 \pm 0.02$ & $0.46 \pm 0.05$ \\
0.50 < $z$ < 0.80 & $0.15 \pm 0.02$ & $1.07 \pm 0.04$ & $0.26 \pm 0.01$ & $1.36 \pm 0.04$ & $0.12 \pm 0.03$ & $0.83 \pm 0.07$ \\
0.80 < $z$ < 1.50 & $0.12 \pm 0.03$ & $1.32 \pm 0.06$ & $0.27 \pm 0.02$ & $1.63 \pm 0.04$ & $0.14 \pm 0.03$ & $1.23 \pm 0.07$ \\
1.50 < $z$ < 2.50 & $0.16 \pm 0.02$ & $1.60 \pm 0.05$ & $0.24 \pm 0.02$ & $1.72 \pm 0.05$ & $0.16 \pm 0.02$ & $1.51 \pm 0.05$ \\
\hline
$z$ & $1 / \gamma_{\mathrm{D15}}$  &
-$\frac{\mathrm{log}_{10} \delta_{\mathrm{D15}}}{\gamma_{\mathrm{D15}}}$ &
$1 / \gamma_{\mathrm{D15}} \langle \mathrm{100 Myr} \rangle$ &
log$_{10} \delta_{\mathrm{D15}} \langle \mathrm{100 Myr} \rangle$ &
$1 / \gamma_{\mathrm{D15}}$*  &
-$\frac{\mathrm{log}_{10} \delta_{\mathrm{D15}}}{\gamma_{\mathrm{D15}}}$* \\
\hline\hline
0.01 < $z$ < 0.25 & $1.07 \pm 0.10$  & $-3.21 \pm 0.05$ & $1.06 \pm 0.06$ & $-3.29 \pm 0.04$ & $2.68 \pm 0.14$ & $-5.52 \pm 0.07$ \\
0.25 < $z$ < 0.50 & $1.13 \pm 0.11$  & $-3.30 \pm 0.05$ & $0.94 \pm 0.04$ & $-3.07 \pm 0.02$ & $2.28 \pm 0.20$ & $-5.61 \pm 0.10$ \\
0.50 < $z$ < 0.80 & $1.06 \pm 0.19$  & $-3.35 \pm 0.17$ & $1.10 \pm 0.03$ & $-3.23 \pm 0.02$ & $2.21 \pm 0.33$ & $-5.99 \pm 0.29$ \\
0.80 < $z$ < 1.50 & $1.00 \pm 0.01$  & $-3.39 \pm 0.15$ & $1.27 \pm 0.03$ & $-3.44 \pm 0.03$ & $2.48 \pm 0.21$ & $-6.53 \pm 0.23$ \\
1.50 < $z$ < 2.50 & $0.96 \pm 0.16$  & $-3.17 \pm 0.24$ & $1.37 \pm 0.04$ & $-3.60 \pm 0.05$ & $1.61 \pm 0.26$ & $-5.18 \pm 0.39$ \\
\hline
\end{tabular}
\label{table:slopes}
\end{table*}

We begin by comparing the predicted relationship between galaxy and BH growth
rates to two recent observational studies using different selection criteria.
Firstly, we explore the \av{SFR} versus BHAR relation for the \textit{AGN
selected} sample presented by \citet{Stanley2015}.  Secondly, we explore the
\av{BHAR} versus SFR relation for the \textit{SFR selected} sample presented by
\citet{Delvecchio2015}. Together they represent two of the largest sample sizes
of their respective selection techniques, spanning multiple epochs. Large
sample sizes such as these are key in overcoming the uncertainties inherent to
low number statistics and in mitigating the potential redshift evolution biases
that could be misinterpreted as an underlying trend.

As mentioned in \cref{sect:individual_galaxies}, SFRs obtained via FIR tracers
most likely probe the recent star formation history of a galaxy, rather than an
instantaneous value. Therefore the comparative SFRs of \eagle galaxies used
in the analysis of \cref{fig:stanley2015,fig:delvecchio2015} are the
time-averaged rate over the 100~Myr preceding the instantaneous BHAR
measurement. However, when performing the equivalent analysis using the
instantaneous values of SFR at the time of the BHAR measurement we find no
difference in the result, attesting to the stability of star formation as a
process.   

\subsubsection{SFR as a function of BHAR for an AGN selected sample}

The study of \citet{Stanley2015} consists of $\approx 2000$ X-ray detected AGN
spanning over three orders of magnitude in luminosity ($10^{42} < $ \LXray$<
10^{45.5}$~erg~s$^{-1}$) for the redshift range $z=0.2 - 2.5$. To compare to
the data we convert their quoted bolometric AGN luminosities (derived from
X-ray luminosities) to BHARs via  

\begin{equation} 
\dot M_{\mathrm{BH}} = \frac{L_{\mathrm{AGN}}}{\epsilon_{r} c^{2}},
\label{eq:lagn}
\end{equation}

\noindent where $c$ is the speed of light and $\epsilon_{r}$ is the radiative
efficiency of the accretion disk, which is assumed to be 0.1
\citep{Shakura1973}. To emulate the sample selection of this study, we choose
\eagle galaxies based on a redshift independent detection limit of $\dot
M_{\mathrm{BH}} = 10^{-3}$~\Msol~yr$^{-1}$, corresponding to \LAGN$\approx
10^{43}$~erg~s$^{-1}$ according to \cref{eq:lagn} and is equivalent to \LXray
$\approx 10^{42}$~erg~s$^{-1}$ using the conversion methods outlined in Section
3.2 of \citet{Stanley2015}.

The results are presented in \cref{fig:stanley2015}, showing the
\textit{mean} SFR as a function of BHAR, represented by a solid black line. We
see that for each redshift range the gradient of the relation remains shallow
(note the same dynamic range is used for both axes), ranging in values $0.1
\leq \gamma_{\mathrm{S15}} \leq 0.2$ (see \cref{table:slopes}) and is in
striking agreement with the \dquotes{remarkably flat} relation reported by
\citet{Stanley2015}.  This is considerably flatter than one would predict for a
linear \M{BH}--\M{bulge,*} relation from a co-evolution of growth, which we
represent as a dashed green line\footnote{Using \cref{eq:slope3} with $\beta =
1000$ \citep{McConnellandMa2013}.} (see \cref{sect:motivation}). The dynamic
range of SFRs is modest, with a scatter of $\approx 1-1.5$~dex for all
redshifts.  The normalisations of \av{SFR} in the three lowest redshift ranges
($0.2 < z < 1.5$) are in good agreement with the observational estimates
(within $\approx 0.1$~dex).  However, the values in the highest redshift range
($1.5 < z < 2.5$) are systematically under predicted by $\approx 0.5$~dex over
and above the recalibration discussed in \cref{sect:sfr_calibration}. We note
that this highest bin is potentially subject to the largest systematic over
estimate ($\approx 0.4$~dex) due to the large fraction ($\approx 80$\%) of
undetected FIR sources (included as upper limits) in the observations
\citep[see][]{Stanley2015}.

\subsubsection{BHAR as a function of SFR for a SFR selected sample}

The study of \citet{Delvecchio2015} consists of $\approx 8600$ star-forming
galaxies detected out to $z=2.5$. The selection limits in SFR are redshift
dependent, corresponding to 0.2, 1.0, 3.0, 8.0 and 25.0 \Msolyr for the five
redshift ranges covered by this study from low to high respectively. We note
that the data points from \citet{Delvecchio2015} are segregated also in stellar
mass, however for simplicity we make no such distinction.

The comparison is shown in \cref{fig:delvecchio2015}, showing the \textit{mean}
BHAR as a function of SFR as the solid black line.  Again, \eagle shows a good
consistency with the observational measurements (shown in red), only over
predicting \av{BHAR} in the lowest redshift range ($0.0 < z < 0.5$).  However,
\citet{Delvecchio2015} mentions that the limited co-moving volume of this study
at low redshift could potentially exclude the most luminous sources.  The
behaviour of the \av{BHAR}--SFR relation is quite different from the
\av{SFR}--BHAR relation seen in \cref{fig:stanley2015}, adhering much closer to
a linear trend. We see, uniformly, gradients close to unity ($1.0 \leq 1 /
\gamma_{\mathrm{D15}} \leq 1.2$, see \cref{table:slopes}) in good agreement
with the linear \M{BH}--\M{bulge,*} relation expected for a co-evolution of
growth, shown as a dashed green line (note again the same dynamic range is used
for both axes).  An additional difference is the spread of values in the
minimisation axis (\BHAR for this figure). The distribution of SFRs in
\cref{fig:delvecchio2015} span a relatively narrow dynamic range, whereas here,
BHARs vary as much as $\approx 4$~dex in the $10-90^{\mathrm{th}}$ percentile
region.  In fact, the dynamic range of BHARs is so large that the small
fraction of galaxies whose values dominate the mean are able to pull it outside
this percentile range entirely in some places, suggesting the median to be a
more suitable statistic to measure this trend. 

Overall the agreement between \eagle and the observations is excellent,
particularly given that no information regarding this relation was considered
during the calibration procedure. The difference in behaviour found empirically
via alternate selection criteria is well reproduced by the simulation. We find,
consistent with the \citet{Hickox2014} model and findings by
\citet{Volonteri2015b}, that \av{SFR}--BHAR for an AGN selected sample exhibits
a relatively flat trend ($\gamma_{\mathrm{S15}} \approx 0.15$), whilst that of
\av{BHAR} with respect to SFR for a SFR selected sample is substantially
steeper and close to unity ($1 / \gamma_{\mathrm{D15}} \approx 1.1$). However,
within the paradigm of a linear \M{BH}--\M{bulge,*} relation created through
co-evolution of growth these results are both predicted to be linear (i.e,
$\gamma \approx 1$ for both, see \cref{sect:motivation}).  Therefore either the
underlying relationship itself is fundamentally non-linear, or a fuller
understanding of the two processes is required. In the next section we continue
to examine potential reasons as to the cause of this difference. 

\subsection{Understanding the BHAR-SFR relationship} \label{sect:understanding}

In this section we explore two potential reasons why the \av{SFR}--BHAR and
\av{BHAR}--SFR trends are not each consistent with a linear relationship. We
examine the hypothesis that (1) growth rates have an underlying linear
connection only on average, which is masked when the unstable growth rate is
observed instantaneously and (2) how selection biases due to the inability to
probe the complete SFR--BHAR plane may play a role.  

\subsubsection{A time-averaged SFR--BHAR connection}
\label{sect:time_av}

\citet{Hickox2014} suggested that an underlying correlation held between a
stable (galactic star formation) and unstable (BH accretion) process \textit{on
average} could be washed out if the unstable property is measured
instantaneously.  That is to say, if one could observe X-ray luminosities of
AGN sources over prolonged periods, the underlying relationship between the two
properties would begin to emerge.  Indeed, with a simple model that assumes SFR
and BHAR are connected on average with a linear scaling over a period of
100~Myr, \citet{Hickox2014} reproduce the empirical behaviours of both the
\av{SFR}--BHAR and \av{BHAR}--SFR relationships whilst retaining a scenario
consistent with a linear co-evolution between galaxies and their central BHs.
While it is not possible to test observationally due to the length of these
timescales, we are able to test this hypothesis using the simulation.

\begin{figure}
\includegraphics[width=\columnwidth]{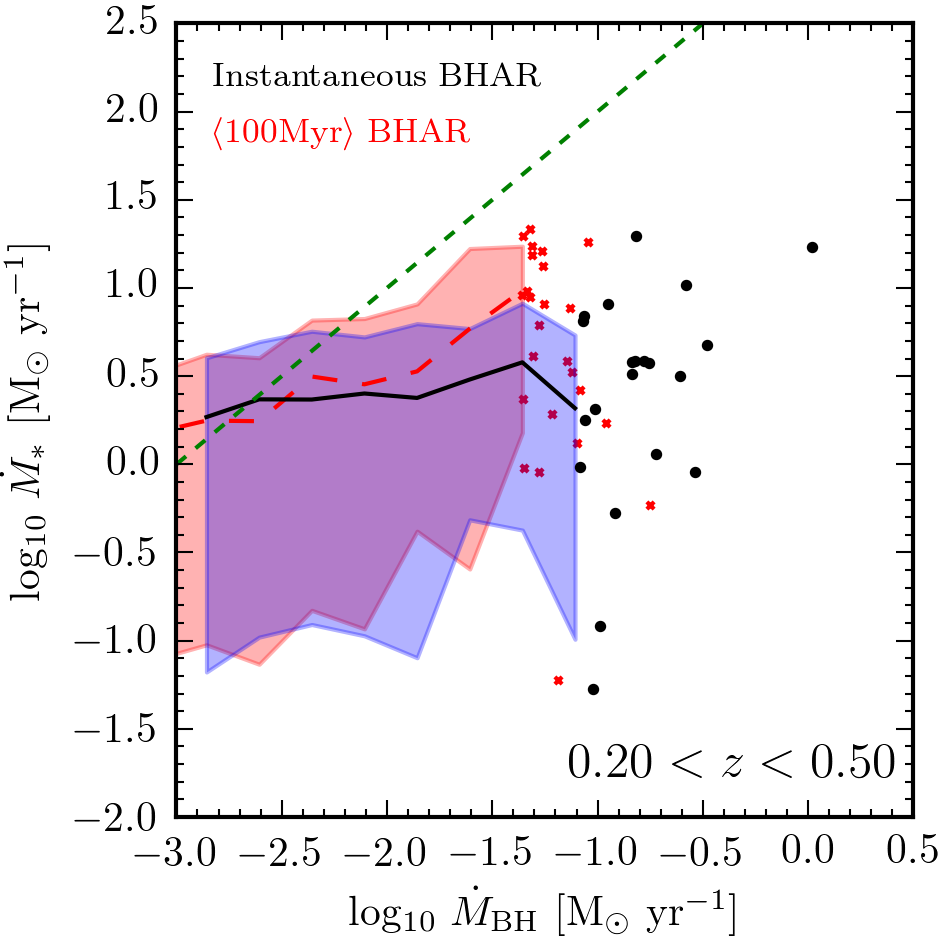}

\caption{\textit{blue region}: A replica of the upper left panel in
\cref{fig:stanley2015} (see caption for description of contents). Here BHARs
are instantaneous and SFRs are the time-averaged rate over the 100~Myr
preceding the BH event. \textit{red region}: We repeat the analysis for the
same central galaxies satisfying the \citet{Stanley2015} selection criteria
(instantaneously, blue region), however now both SFR and BHAR are time-averaged
over the same 100~Myr period.  Fits to the time-averaged mean relations are
shown in \cref{table:slopes} (denoted with $\langle 100~ \mathrm{Myr}
\rangle$). We find that even when both growth rates are time-averaged over
100~Myr, an AGN selected sample does not revert to a linear relationship
between \SFR and \BHAR.}

\label{fig:avandinst} \end{figure}

\cref{fig:avandinst} is similar to the upper left panel of
\cref{fig:stanley2015}. The region in blue, with the black solid line, shows
the original analysis of the \av{SFR}--BHAR relation for the redshift range
$0.2 < z < 0.5$ using instantaneous BHARs and SFRs that are time-averaged 100~Myr
before the BH event. Overlaid in red, with the mean represented by a dashed
line, is the same selection of galaxies (i.e, instantaneous \BHAR $>
10^{-3}$\Msolyr) with each growth rate now time-averaged over 100~Myr.
Interestingly, although the high BHARs shift systematically to lower values on
average\footnote{The shift ($\approx$0.5~dex) to lower values in BHAR when
averaging over 100~Myr arises due to the most luminous \squotes{detections}
commonly residing in peaks of the accretion rate history.}, both the dynamic
range and slope of the mean remain broadly consistent with their instantaneous
equivalents ($\gamma_{\mathrm{S15}} \rightarrow \gamma_{\mathrm{S15} \langle
100~\mathrm{Myr} \rangle} = 0.2 \rightarrow 0.3$, see \cref{table:slopes}).
This behaviour remains for each redshift range explored by \citet{Stanley2015}
(see \cref{table:slopes}).  An alternate approach is to select galaxies in
excess of \BHAR $= 10^{-3}$\Msolyr on average (rather than instantaneously as
done above) or indeed to prolong the averaging timescale to $> 100$~Myrs.
However in each case, and for all redshift intervals, the gradient values
remain well below unity ($0.30 < \gamma_{\mathrm{S15} \langle 100~\mathrm{Myr}
\rangle} < 0.55$).  This leads us to conclude that the average galaxy and BH
growth rates for an AGN selected sample do not harbour an underlying global
linear relationship. 

\subsubsection{Sampling different regions of the entire SFR-BHAR plane}
\label{sect:sfr_vs_bhar_plane}

\begin{figure}
\includegraphics[width=\columnwidth]{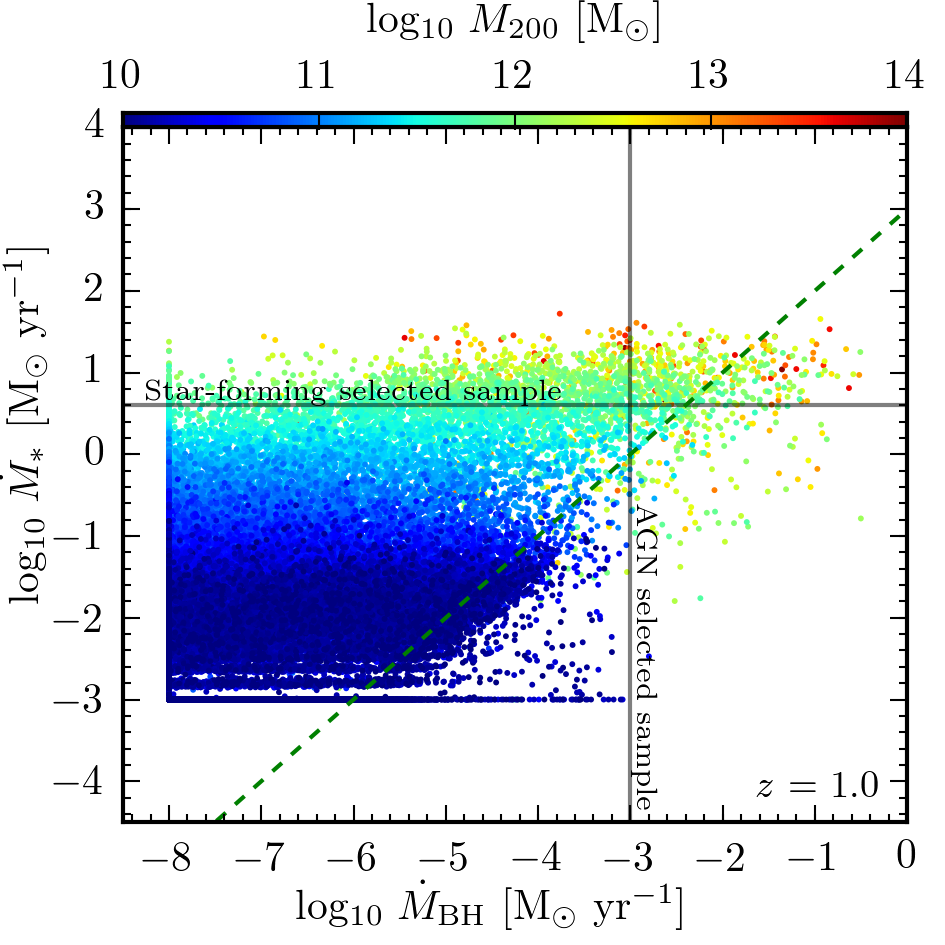}

\caption{\SFR-\BHAR (SFR-BHAR) relation for central galaxies at $z=1$. Growth
rates are \textit{instantaneous} and coloured by the mass of the halo
(\M{200}). Values that are below $10^{-8}$~\Msolyr for \BHAR and below
$10^{-3}$~\Msolyr for \SFR are clipped to these limits. The approximate flux
limits of \citet{Stanley2015} and \citet{Delvecchio2015} investigated in
\cref{sect:observations} are shown as vertical and horizontal solid lines
respectively, highlighting the different regions of the full distribution that
these surveys are able to probe. The dashed green line indicates the linear
relation \BHAR/\SFR = $10^{-3}$.}

\label{fig:sfr_vs_bhar_z1}
\end{figure}

\begin{figure}
\includegraphics[width=\columnwidth]{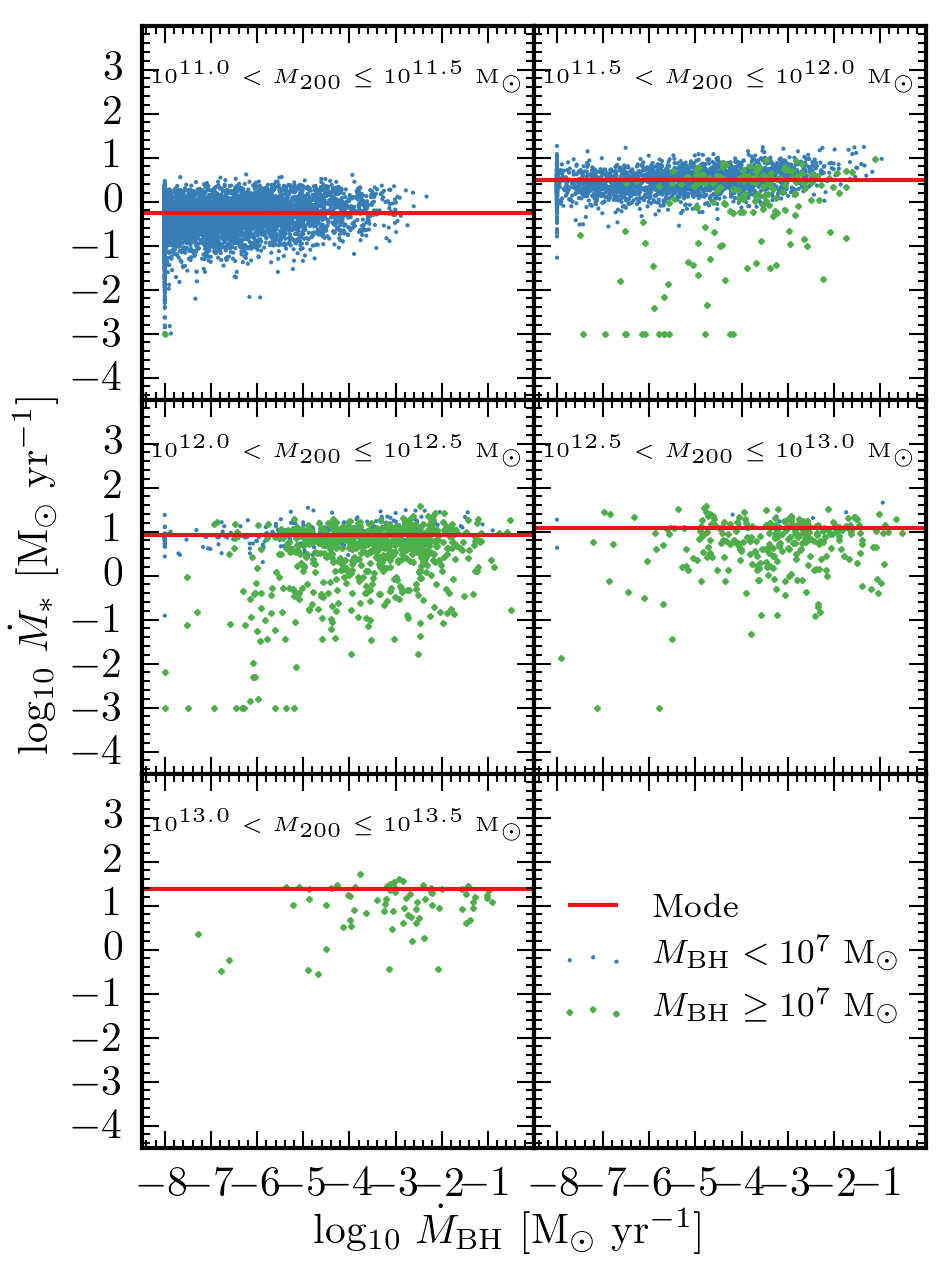}

\caption{Growth rates at $z=1$ (from \cref{fig:sfr_vs_bhar_z1}) for central
galaxies separated into five continuous halo mass ranges. Points are coloured
green if a galaxy hosts a massive BH (\M{BH} $\geq 10^{7}$\Msol) and blue
otherwise. The characteristic SFR for a given halo mass bin (classified as the
mode of the distribution) is shown as a horizontal solid red line. Each range
of halo mass yields a relatively narrow distribution of SFRs ($1-2$~dex) and
much wider distribution of BHARs (up to 8~dex).  Larger halos are associated
with larger characteristic SFRs and have a higher fraction of BHARs $>
10^{-4}$\Msolyr. Galaxies harbouring SFRs far below the characteristic rate all
contain massive BHs (green points) and have likely recently undergone a violent
episode of AGN feedback reducing the current star-forming capability of the
system.}

\label{fig:sfr_vs_bhar_byhalo} \end{figure}

As observational surveys are subject to various flux limitations, they can only
sample particular regions of the full SFR--BHAR plane.  If the global
underlying relation is linear, and each property exhibits a moderate scatter,
each sub sample should also return a linear result.  However, as the findings
of previous sections do not to support an underlying linear relation, and
because the scatter is large, it is important to investigate the effect of this
sampling. 

\cref{fig:sfr_vs_bhar_z1} shows the complete SFR--BHAR plane for all central
galaxies at $z=1$. In order to eliminate any potential bias incurred via
redshift evolution in either growth rate we consider a discrete redshift rather
than the continuous ranges of \cref{sect:observations}. Each data point
represents the \textit{instantaneous} state of a single galaxy and its central
BH, coloured by the halo mass.  Values that are below $10^{-8}$~\Msolyr for
\BHAR and $10^{-3}$~\Msolyr for \SFR are treated as \squotes{zero} and are
clipped to these values. The approximate flux limit of the AGN selected sample
by \citet{Stanley2015} is shown as a solid vertical line and the approximate
flux limit of the SFR selected sample by \citet{Delvecchio2015} is shown as a
horizontal solid line.

It is further apparent that the global relationship between SFR and BHAR is not
simply linear (reference with the dashed green line). Instead, a complicated
relationship arises due to an amalgamation of three distinct behaviours of BH
growth dependent on the mass of the host dark matter halo (see next section).
It is therefore crucial to consider the particular region sampled before
arriving at a particular conclusion.  AGN selected samples, such as that of
\citet{Stanley2015}, currently probe a relatively limited region at the tip of
the SFR--BHAR plane.  With the exception of a few sources with rates \SFR $\ll
1$\Msolyr, galaxies satisfying this selection criteria are distributed over a
relatively narrow range of SFRs. As such, each bin of BHAR yields a very
similar value of \av{SFR}, creating an approximately flat trend.  SFR selected
samples, such as that of \citet{Delvecchio2015}, sample a not too dissimilar
distribution of SFRs (this time due to the flux limit), however, the
distribution of BHARs is much wider.  This in turn yields a steeper relation.
We note, that whilst the mean SFR provides a good proxy of the median SFR for
an AGN selected study (compare columns 3 and 6 of \cref{table:slopes}), the
mean BHAR for a SFR selected study is not a good proxy of the median value due
to the distribution of BHARs having such a large scatter (compare columns 8 and
10 of \cref{table:slopes}).  Although only the results from $z=1$ have been
shown here, when investigated we find the results remain true independent of
redshift.

We therefore conclude that the different behaviour found for the \av{SFR}--BHAR
and \av{BHAR}--SFR relations recovered by observational studies is due to
sampling considerably different regions of the full (not universally linear)
SFR--BHAR plane. We now continue to investigate the nature of this relationship
in the \eagle simulation and its evolution through time.

\subsection{The connection to the host dark matter halo}
\label{sect:to_halo}

The relationship between SFR and BHAR seen in \cref{fig:sfr_vs_bhar_z1} is
complicated, seemingly not adhering to a simple universal trend.  However,
there is evidence that each property has a link with the mass of the host dark
matter halo, highlighted in the change of the data point colours, which
transition smoothly from blue to red with increasing SFR, and systematically
shift rightward in BHAR (with a large scatter) at high halo mass.    

To examine this in more depth we sub categorize the $z=1$ central galaxies into
five continuous ranges of halo mass, showing the growth rates in
\cref{fig:sfr_vs_bhar_byhalo}.  Here we find, in fact, that the global make up
of the SFR--BHAR plane in \cref{fig:sfr_vs_bhar_z1} is resolved into a
collection of two dimensional strips, wide in their dynamic range of BHAR
($\approx 4-6$~dex) yet generally much more compact in their SFR ($\approx
1-2$~dex).  Each strip hosts a characteristic value of SFR (defined as the mode
of the distribution, shown as a horizontal solid red line) that continuously
increases with increasing halo mass. This is in line with the \squotes{star
forming main sequence}, where galaxies of increased stellar mass are seen to
host larger SFRs \citep[e.g,][]{Elbaz2007}. Interestingly, the rate of change
with \M{200} for this characteristic SFR does not remain constant, initially
increasing by $\Delta$\SFR $\approx 1$ dex between $10^{11.0} <$ \M{200} $ \leq
10^{12.0}$\Msol and reducing to almost zero in the regime $10^{12.5} < $
\M{200} $ \leq 10^{13.5}$\Msol. This is potential evidence that SFRs in massive
systems are not keeping pace with the increasing baryonic inflow rates for
increasing halo mass at fixed redshift \citep[e.g,][]{Correa2015}. BHARs show a
less continuous behaviour however, broadly categorized by two rudimentary
states: BHs residing in haloes below $\approx 10^{11.5}$\Msol are typically
accreting at a \squotes{low} rate (\BHAR $\ll 10^{-4}$~\Msolyr); BHs residing
in haloes more massive than $\approx 10^{12.5}$\Msol tend to be accreting at a
\squotes{high} rate (\BHAR $> 10^{-4}$~\Msolyr). The fraction of galaxies with
\BHAR $\geq 10^{-4}$\Msolyr for a given halo mass bin is $\approx$ 3\%, 21\%,
55\%, 70\% and 78\% from $10^{11.0} <$ \M{200} $\leq 10^{11.5}$\Msol to
$10^{13.0} <$ \M{200} $\leq 10^{13.5}$\Msol respectively.  Those in haloes
between the mass range $10^{11.5} \sim 10^{12.5}$\Msol are in an intermediate
state. 

A fraction of galaxies hosted by haloes with \M{200} $\gtrsim 10^{11.5}$\Msol
harbour extremely low, or even zero SFRs. As all of these galaxies host massive
BHs (\M{BH} $\geq 10^{7}$\Msol, green dots), we are most likely seeing the
effect of recent episodes of violent AGN feedback that have severely reduced
the current star-forming capabilities of these systems.  The cause, prevalence
and impact of these feedback events will be the subject of a future paper.

\begin{figure}
\includegraphics[width=\columnwidth]{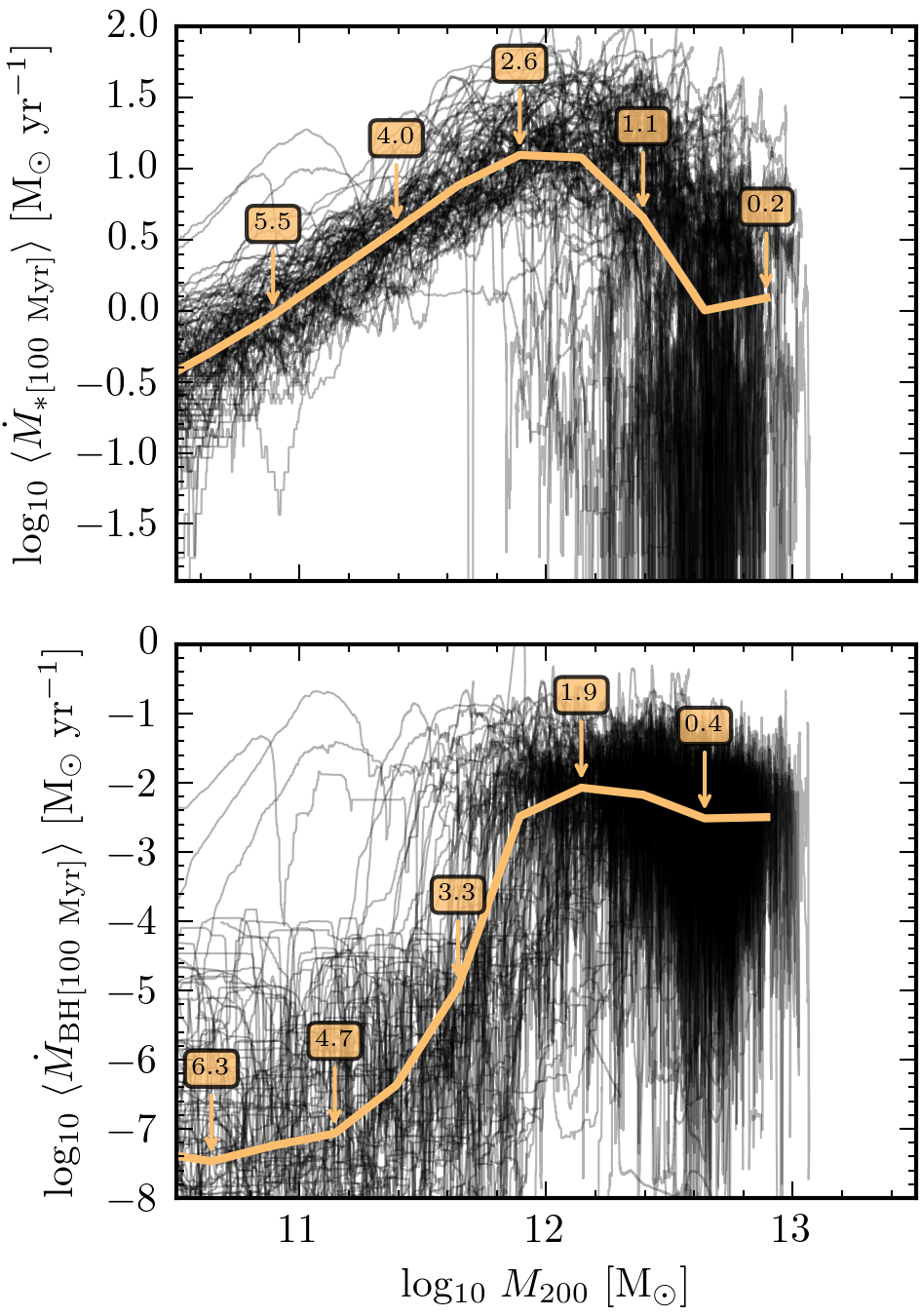}

\caption{The evolutionary history of SFR (top panel) and BHAR (bottom panel) as
a function of halo mass for all galaxies that come to reside in the $10^{12.5}
< M_{\mathrm{200}} < 10^{13.0}$\Msol, $10^{8.0} < M_{\mathrm{BH}} <
10^{8.5}$\Msol two-dimensional bin of \cref{fig:bhm_vs_hm} at $z=0$ (outlined
in yellow). Each black line is an individual history. The orange line shows the
median trend, annotated with the median redshift at which these galaxies were
hosted in haloes of that mass.  For each panel, growth rates are time-averaged
over 100~Myr as to overcome the noise induced when considering instantaneous
rates.  We see very different evolutionary behaviour for SFR and BHAR as the
halo grows. SFRs initially rise and then decline, centred around \M{200} $\sim
10^{12}$\Msol. BHARs similarly transition from a low to high rate around this
halo mass.}

\label{fig:one_square} \end{figure}

\begin{figure}
\includegraphics[width=\columnwidth]{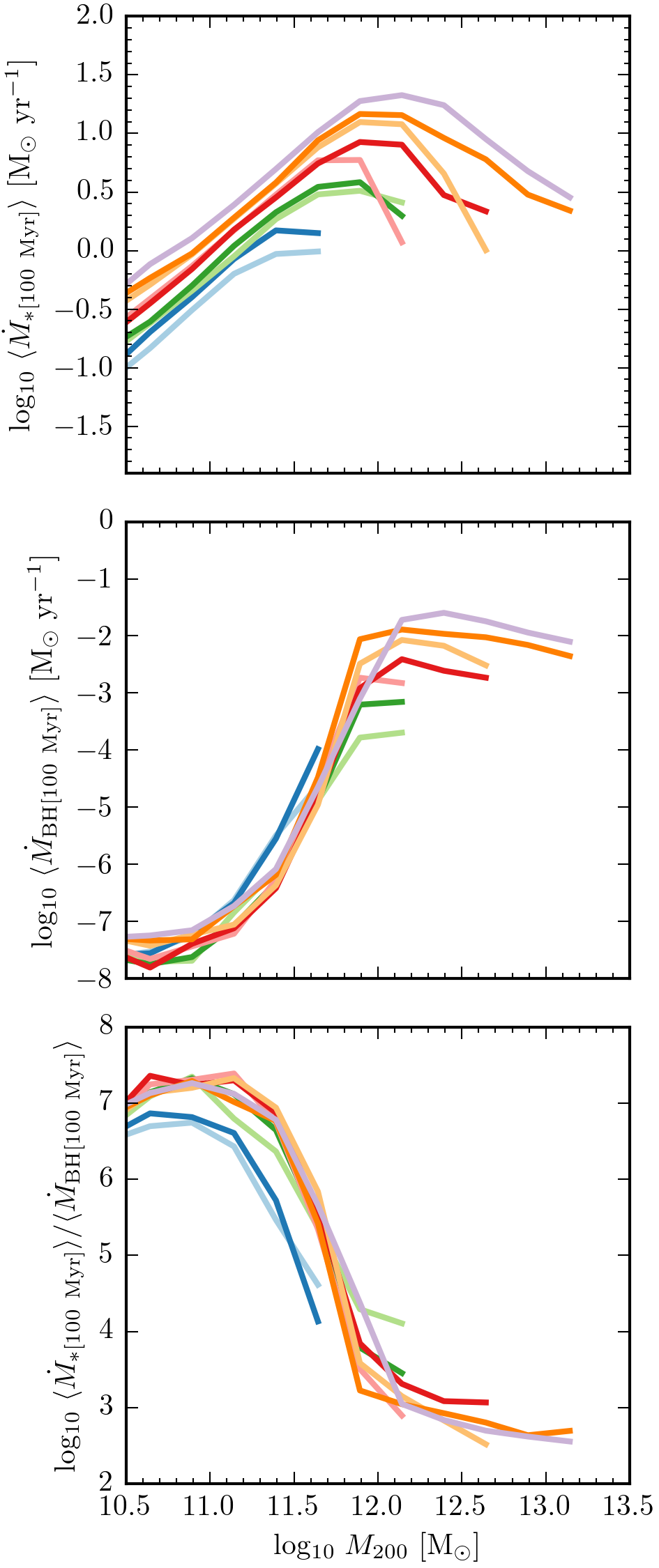}

\caption{\textit{top two panels}: A continuation of the analysis in
\cref{fig:one_square} to each of the nine chosen two-dimensional bins of
\cref{fig:bhm_vs_hm}. The lines are the median track, with the colour
corresponding to the outline in \cref{fig:bhm_vs_hm}. Regardless of where a
galaxy is located on the \M{BH}--\M{200} plane at the present day, both the
galaxy and its central BH evolve similarly, though different from each other.
The change in normalisations between the histories is due to the declining
baryonic inflow rates with decreasing redshift for a fixed halo mass.
\textit{bottom panel}: The median ratio between the SFR and BHAR from the two
panels above. SFRs are initially dominant by many orders of magnitude in low
mass haloes (\M{200} $\lesssim 10^{11.5}$\Msol), coming to plateau at an
approximately constant value of $\sim 10^{3}$ in high mass haloes (\M{200}
$\gtrsim 10^{12.5}$\Msol).}

\label{fig:avHistory_vs_hm}
\end{figure}

We now investigate if the growth rate to halo connection evolves.  To do this
we return to the \M{BH}--\M{200} relation shown for $z=0$ in
\cref{fig:bhm_vs_hm}. The population is sub divided into two-dimensional bins,
0.5~dex on a side and outlined as squares.  Here we investigate nine bins that
lie along the median track through a continuous range spanning $10^{11.5} <$
\M{200} $< 10^{13.5}$\Msol in halo mass and $10^{6.0} <$ \M{BH} $<
10^{9.0}$\Msol in BH mass, each outlined with a unique colour to reference
their histories in
\cref{fig:one_square,fig:avHistory_vs_hm,fig:sfr_vs_bhar_av}.

\cref{fig:one_square} shows the evolution of the time-averaged SFR (top panel)
and time-averaged BHAR (bottom panel) as a function of halo mass for all
galaxies that come to reside in one of these two-dimensional bins at the
present day (each solid black line is an individual history). We time-average
both SFR and BHAR over 100~Myr in order to remove the inherent noise when
considering instantaneous growth rates, and unveil the average trend.  To
eliminate galaxies that were previously classified as satellites of a more
massive halo, we only consider central galaxies that have evolved monotonically
in their halo mass (this excludes only $\approx 1\%$ of the $z=0$ centrals
population).  We see, that although individual histories can be quite
different, on average galaxies and their central BHs do follow a well defined
path.  The median SFR and BHAR of this population subset for a given halo mass
are over-plotted in yellow, annotated by the median redshift at which they were
hosted by haloes of that particular mass.  As expected, an increasing halo mass
corresponds to a decreasing redshift.

There is a striking difference in behaviour seen between the two growth rates
as the halo grows.  Initially the SFR increases steadily with halo mass. As the
halo grows more massive than $\approx 10^{12}$\Msol the SFRs begin to fluctuate
between high and low values, yet overall there is a gradual decline of the
median trend after this mass.  Similarly, BHARs also change their behaviour
around $\approx 10^{12}$\Msol, rapidly transitioning from a low (\BHAR $\ll
10^{-4}$\Msolyr) to high (\BHAR $> 10^{-4}$\Msolyr) rate. As with SFRs, BHARs
decline a similar amount after the halo mass $\approx 10^{12}$\Msol (note the
many orders of magnitude difference in the scale of the growth rate axis
between the two panels).  We interpret therefore, given that the decline of SFR
coincides with the peak of the rapid increase in BHAR, that AGN feedback is
impeding the continued rise of SFRs in the most massive systems (see
\cref{fig:variability} for an individual example of SFR reduction after the
peak AGN activity at lookback time $\approx 12$).  We note that the decrease in
halo mass accretion rate with declining redshift and the dependence of halo
cooling rates on halo mass will play \textit{additional} roles in shaping these
histories. However, given the severity of the SFR reduction seen immediately
after the BHAR peak, AGN feedback appears to be a dominant factor in hindering
further galaxy growth.

\cref{fig:avHistory_vs_hm} extends this analysis to each of the highlighted
two-dimensional bins in \cref{fig:bhm_vs_hm}, now showing only the median lines
for clarity. Remarkably, the evolutionary behaviour is similar regardless of
the final position in the \M{BH}--\M{200} plane.  The normalisation of each
history is set by the evolving baryonic inflow rate at fixed halo mass. As this
rate decreases with redshift \citep[e.g,][]{Correa2015}, so does the
normalisation of both the SFR and BHAR seen here (as each population reaches a
particular halo mass at different times). We include also in the bottom panel
of \cref{fig:avHistory_vs_hm} the median ratio between SFR and BHAR shown in
the two panels above. This shows that galaxy growth is dominant over BH growth
by many orders of magnitude in low mass haloes (\M{200} $\lesssim
10^{11.5}$\Msol). As BHARs settle to their \squotes{high} rate in haloes of a
mass above \M{200} $\sim 10^{12}$, the ratio between SFR and BHAR plateaus to
an approximately constant value of $\sim 10^{3}$.  Note the trends of both
\cref{fig:one_square,fig:avHistory_vs_hm} are not directly observable as they
rely on median time-averaged growth rates in both SFR and BHAR of 100~Myr
whilst also being binned by halo mass. 

\section{Discussion}
\label{sect:discussion}

\begin{figure}
\includegraphics[width=\columnwidth]{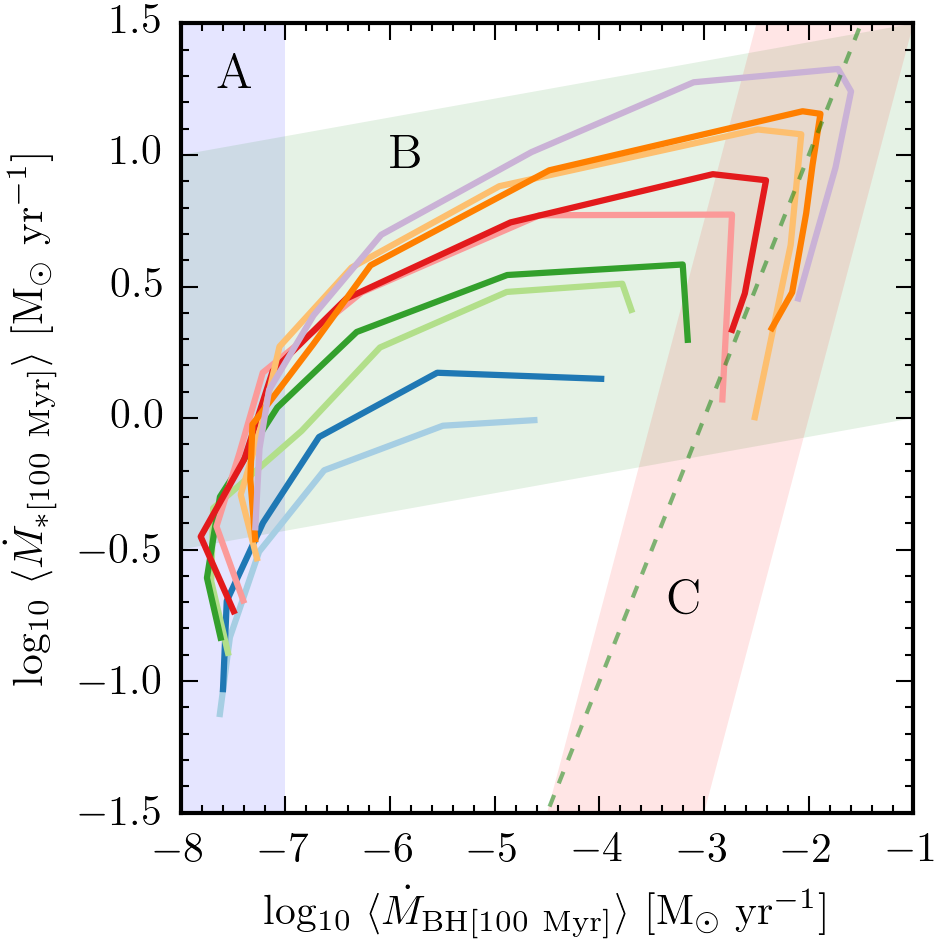}

\caption{Each line shown here equates the median trends from the top two panels
of \cref{fig:avHistory_vs_hm} to give the 100~Myr average SFR as a function of
the 100~Myr average BHAR (in equal spacings of halo mass). Region \emph{A}
(shaded blue) corresponds to galaxies hosted by haloes with \M{200}
$\lesssim$\M{crit}. Galaxies in this regime increase their SFR with increasing
halo mass, while BHARs remain negligible on average. As haloes reach
$\sim$\M{crit} in region \emph{B} (shaded green), SFRs continue to rise,
however the BH growth increases by many orders of magnitude over this narrow
halo mass range. For haloes in excess of $\gtrsim$ \M{crit} shown in region
\emph{C} (shaded red), we see a reduction for both SFR and BHAR on average,
yielding a approximately constant scaling between the two growth rates (compare
to dashed green line which shows the linear relation \BHAR/\SFR = $10^{-3}$).}

\label{fig:sfr_vs_bhar_av} \end{figure}

Throughout this investigation we have consistently found no evidence supporting
a simple underlying relationship between the rate of a galaxy's star formation
and the accretion rate of its central BH. Instead, a mutual dependence of each
property upon the mass of the host halo yields a more complex connection. It is
interesting to examine, then, how the relation between the SFR and BHAR evolves
for individual objects. In the following discussion, we will provide a physical
interpretation based on the \citet{Bower2017} (hereafter B16) model for BH
growth \citep[for a similar interpretation on the importance of SN feedback to
BH growth see][]{Dubois2015,Habouzit2016}.  However, we stress the simulation
results are themselves independent of any physical interpretation.  

\cref{fig:sfr_vs_bhar_av} equates the median trends of the SFR and BHAR
histories shown in \cref{fig:avHistory_vs_hm}. This specifies the 100~Myr
average SFR as a function of the 100~Myr average BHAR in equal spacings of halo
mass.  Three distinct trends between SFR and BHAR emerge as the halo evolves:
the \emph{stellar feedback regulated} phase (shaded blue), the \emph{non-linear
BH growth} phase (shaded green) and the \emph{AGN feedback regulated} phase
(shaded red).

\begin{itemize}

\item \emph{Region A - The stellar feedback regulated phase}: From the time of
their seeding until they are hosted by haloes of mass \M{200} $\sim
10^{11.5}$\Msol the BH accretion rates are negligible (\BHAR $\leq
10^{-6}$\Msolyr on average).  By contrast, SFRs increase steadily with halo
mass. This behaviour produces the uncorrelated (yet causally connected)
$\sim$vertical trend in region \emph{A}, creating an imbalance of growth within
these systems. As a result, BHs remain close to their seed mass whilst the
halo/galaxy continues to grow around them (see the low-mass region of
\cref{fig:bhm_vs_hm}).

B16 interpret galaxies in this regime as being in a state of regulatory
equilibrium. Energy injected by stars heats the ISM within the stellar vicinity,
ejecting it, and causing it to rise buoyantly in the halo. This in turn creates
an outflow of material balancing the freshly sourced fuel from the cosmic web,
and as such prevents large gas densities from building up within the inner
regions of these low-mass galaxies.  Such low densities, coupled with the
relatively low mass BHs living within these galaxies (BHAR $\propto
M_{\mathrm{BH}}^{2}$), ensures that BHs fail to grow substantially. 

\item \emph{Region B - The non-linear BH growth phase}: Both galaxies and BHs
grow through the halo mass range \M{200} $\sim 10^{11.5} - 10^{12.0}$\Msol.
However, whereas the SFRs continue to increase steadily with increasing halo
mass, BHs rapidly transition to a non-linear phase of growth. This creates a
highly non-linear \textit{indirect} correlation between SFR and BHAR, connected
through the host halo mass.  

The physical interpretation posited by B16 is that haloes that grow to the
transition mass, \M{crit}, have become sufficiently massive to stall the
regulatory outflow. Due to (what is now) the halos' hot coronae, heated gas
ejected by stellar feedback loses the capability to rise buoyantly and
therefore returns to the galaxy centre.  Densities in the central regions of
the galaxy are no longer kept low and a \squotes{switch} to non-linear BH
growth is triggered.  

\item \emph{Region C - The AGN feedback regulated phase}: For haloes with
masses above \M{200} $\sim 10^{12}$\Msol SFRs and BHARs both decline on
average, correlated with an approximately linear trend (compare to green dashed
line, see also bottom panel of \cref{fig:avHistory_vs_hm}). 

B16 argue that BHs in these haloes have become sufficiently massive (through
their rapid non-linear growth) to efficiently regulate the gas inflow onto the
galaxy themselves via AGN feedback.  This again creates an equilibrium state,
for which a fluctuating low level of (specific) BH accretion is maintained,
keeping the outer halo hot and evaporating much of the new cold material trying
to enter the system from the intergalactic medium.

\end{itemize}

Galaxies and their central BHs within the \eagle simulation transition through
multiple stages of growth as their host dark matter halo evolves, creating
three distinct behaviours between SFR and BHAR. This is a stark contrast to a
simple model where SFR and BHAR correlate globally via a linear relation, on
average and for all halo masses. Whilst the underlying trend is only revealed
when each growth rate is time-averaged (given the inherent noise of
instantaneous growth rates), we only find an approximately linear correlation
for the most massive systems (\M{200} $\gtrsim 10^{12.5}$\Msol).

In this paper we have emphasised the role of the halo and how its interaction
with both SFR and BHAR shapes the growth rate relationship.  However,
additional factors may also contribute to the form this relationship takes. For
example, \citet{Volonteri2015b} find using a suite of isolated merger
simulations at fixed halo mass, that alternate behaviours between SFR and BHAR
before, during and after the merger proper collectively contribute to form a
complex two-dimensional plane.  Additionally, \citet{Pontzen2017} reveal the
particular importance differing merger histories can have on significantly
altering the growth rate history of both that of the galaxy and the central BH.
However, the global influence of mergers upon galaxy and BH growth rates in a
full cosmological context remains open for debate, and will be the subject of a
future paper.

\section{Conclusions}
\label{sect:conclusions}

We have investigated the relationship between the galaxy star formation rate
(SFR) and the black hole accretion rate (BHAR) of the central black hole (BH)
using the \eagle cosmological hydrodynamical simulation. Our main conclusions
are as follows: 

\begin{itemize}

\item We compared \eagle predictions to two recent observational studies in
\cref{fig:stanley2015,fig:delvecchio2015}. The simulation reproduces both the
flat trend of the mean SFR (\av{SFR}) as a function of BHAR found in the AGN
selected study of \citet{Stanley2015} and the approximately linear trend of the
mean BHAR (\av{BHAR}) as a function of SFR found in the SFR selected study of
\citet{Delvecchio2015}. 

\item There is a moderate difference in the \av{SFR}--BHAR relationship when
time-averaging each growth rate over a 100~Myr period for an AGN selected study
(\cref{fig:avandinst}).  However, this change was not found to be sufficient as
to revert the trend to an underlying linear relationship as has been proposed by
previous theoretical studies.

\item Examining the complete $z=1$ SFR--BHAR plane in
\cref{fig:sfr_vs_bhar_z1}, we found no evidence for a simple universal global
relationship between the two instantaneous growth rates. The difference between
the trends found for the \av{SFR}--BHAR and \av{BHAR}--SFR relations from AGN
and SFR selections respectively, is due to sampling different regions of this
complex plane. The complexity of this plane results from both the rate of
galactic star formation and the accretion rate of the central BH holding an
evolving connection to the host dark matter halo
(\cref{fig:sfr_vs_bhar_byhalo}).

\item For a discrete redshift, the characteristic SFR of a halo increases
smoothly with increasing halo mass (\cref{fig:sfr_vs_bhar_byhalo}). BHs in
haloes of mass \M{200} $\lesssim 10^{11.5}$\Msol accrete at a \squotes{low}
rate (\BHAR $< 10^{-4}$\Msolyr).  They then transition through haloes of mass
$10^{11.5} \sim 10^{12.5}$\Msol to a \squotes{high} rate (\BHAR $>
10^{-4}$\Msolyr) in haloes of mass \M{200} $\gtrsim 10^{12.5}$\Msol. However,
the scatter in the BHAR at fixed halo mass is very large (up to $\sim 6$~dex).
Galaxies with SFRs far below the characteristic SFR all contain massive BHs
(\M{BH} $\geq 10^{7}$\Msol).  

\item The median evolutionary trend for a galaxy's SFR and the accretion rate
of its central BH, averaged over 100~Myr, are insensitive to the final
properties of the system (\cref{fig:avHistory_vs_hm}). By equating these trends
together we found that the 100~Myr average SFR as a function of the 100~Myr
average BHAR can be split into three regimes, separated by the halo mass
(\cref{fig:sfr_vs_bhar_av}). BHs hosted by haloes below the characteristic
transition mass, \M{crit} \citep[][\M{200} $\sim 10^{12}$\Msol]{Bower2017},
fail to grow effectively, yet the galaxy continues to grow with the halo. Once
the halo reaches \M{crit} there is a non-linear \squotes{switch} of BH growth
that rapidly builds the mass of the BH. In the most massive haloes (\M{200} >
\M{crit}) both SFR and BHAR decline on average, with a roughly constant scaling
of SFR/BHAR $\sim 10^{3}$.

\end{itemize}

\section*{Acknowledgements}

We thank the referee Marta Volonteri for her useful comments, also to Flora
Stanley and Ivan Delvecchio for their data and useful conversations. 

This work was supported by the Science and Technology Facilities Council (grant
number ST/F001166/1 and ST/L00075X/1); European Research Council (grant numbers
GA 267291 \dquotes{Cosmiway} and GA 278594 \dquotes{GasAroundGalaxies}) and by
the Interuniversity Attraction Poles Programme initiated by the Belgian Science
Policy Office (AP P7/08 CHARM). RAC is a Royal Society University Research
Fellow.

This work used the DiRAC Data Centric system at Durham University, operated by
the Institute for Computational Cosmology on behalf of the STFC DiRAC HPC
Facility (\url{http://www.dirac.ac.uk}). This equipment was funded by BIS
National E-infrastructure capital grant ST/K00042X/1, STFC capital grant
ST/H008519/1, and STFC DiRAC Operations grant ST/K003267/1 and Durham
University. DiRAC is part of the National E-Infrastructure. We acknowledge
PRACE for awarding us access to the Curie machine based in France at TGCC, CEA,
Bruy\`{e}res-le-Ch\^{a}tel.

This work was supported by the Netherlands Organisation for Scientific Research
(NWO), through VICI grant 639.043.409, and the European Research Council under
the European Union's Seventh Framework Programme (FP7/2007- 2013) / ERC Grant
agreement 278594-GasAroundGalaxies.

\bibliographystyle{mnras}
\bibliography{mybibfile}

\appendix

\section{Predictions from the integrated quantities}
\label{sect:motivation}

The major motivation for linking the growth of galaxies to the growth of their
central BH has arisen empirically from the strong correlations seen in their
integrated properties (the primary example being the tight \M{BH}--\M{*,bulge}
relation).  We therefore require an evolutionary model that suitably fits this
end point.  In the simplest case, where BHs and their host galaxy grow in
concert (or co-evolve), the relational form between their growth rates can be
easily predicted.  Given a functional form of the \M{BH}--\M{bulge,*} relation
described via

\begin{equation} \mathrm{log}_{10} M_{\mathrm{BH}} = \alpha \mathrm{log}_{10}
M_{\mathrm{bulge,*}} + \mathrm{log}_{10} \beta, \label{eq:slope1}
\end{equation}

\noindent where $\alpha$ is the gradient of the slope and $\mathrm{log}_{10}
\beta$ is the intercept, the predicted relation between the growth rates is
simply found by differentiating with respect to time, i.e,

\begin{equation} \mathrm{log}_{10} \dot M_{\mathrm{BH}} = \mathrm{log}_{10}
\dot M_{\mathrm{bulge,*}} + \mathrm{log}_{10} \beta + \mathrm{log}_{10} \alpha
+ (\alpha -1) \mathrm{log}_{10} M_{\mathrm{bulge,*}} .  \label{eq:slope2}
\end{equation}

\noindent In the trivial case where $\alpha = 1$ (i.e, a linear relation)
this reduces to

\begin{equation} \mathrm{log}_{10} \dot M_{\mathrm{BH}} = \mathrm{log}_{10}
\dot M_{\mathrm{*}} + \mathrm{log}_{10} \beta.  \label{eq:slope3}
\end{equation}

\noindent Within this scenario, growth rates are directly proportional to one
another scaled by the intercept, $\beta$, of the \M{BH}--\M{bulge,*} relation.
Therefore, if the functional form between the growth rates is described via

\begin{equation} \mathrm{log}_{10} \dot M_{\mathrm{BH}} = \gamma
\mathrm{log}_{10} \dot M_{*} + \mathrm{log}_{10} \delta, \label{eq:slope4}
\end{equation}

\noindent where again $\gamma$ and $\delta$ are the slope and intercept values,
we would expect $\delta = \beta$ and $\gamma = 1$ in the case where $\alpha =
1$.

Throughout this study we will test the hypothesis that there exists a broadly
linear co-evolution between galaxies and their central BHs, a plausible
scenario fitting the empirical \M{BH}--\M{*,bulge} relation. We refer
throughout to $\alpha$ as the gradient of the slope between the integrated
properties (\M{BH}--\M{*/bulge,*}) and to $\gamma$ as the slope between each
growth rate (\BHAR--\SFR), both in log space.

\bsp	
\label{lastpage}

\end{document}